\documentclass[cits]{PoS-pdf}
\pdfoutput=1
\usepackage{amsmath}
\usepackage{amssymb}            
\usepackage{mathtools}          
\usepackage{algorithmic}        
\algsetup{indent=2em}

\title{New developments in PJFry}

\ShortTitle{New developments in PJFry}

\author{Jochem Fleischer\\
        Fakult\"at f\"ur Physik, Universit\"at Bielefeld, Universit\"atsstr. 25,  33615
        Bielefeld, Germany
        \\
        E-mail: \email{Fleischer@physik.uni-bielefeld.de}
}
\author{Tord Riemann\\
        Deutsches Elektronen-Synchrotron DESY, Platanenallee 6, 15738 Zeuthen, Germany
        \\
        E-mail: \email{Tord.Riemann@desy.de}
}
\author{\speaker{Valery Yundin}%
        \\
        Niels Bohr International Academy and Discovery Center, The Niels Bohr Institute,\\
        University of Copenhagen, Blegdamsvej 17, DK-2100 Copenhagen, Denmark\\
        E-mail: \email{yundin@nbi.dk}}

\abstract{
We report on recent progress in numerical evaluation of one loop tensor integrals.
A public C++ package PJFry~\cite{Yundin:phd,AlcarazMaestre:2012vp} implementing algorithms from Ref.~\cite{Fleischer:2010sq} and its
extension to hexagons up to rank~6 are presented.
}

\FullConference{Loops and Legs in Quantum Field Theory - 11th DESY Workshop on Elementary Particle Physics\\
                 April 15-20, 2012\\
                 Wernigerode, Germany}

\newcommand{\mbinom}[2]{\biggl(\begin{matrix}#1\\#2\end{matrix}\biggr)}  

\begin{document}

\section{Introduction}
The stable numerical evaluation of tensor integrals is one of the central ingredients of
one loop Feynman diagram calculations. It can also be efficiently applied in
modern techniques like tensorial reconstruction at integrand level \cite{Heinrich:2010ax}
or the Open Loops method \cite{Cascioli:2011va}.

The classic Passarino-Veltman reduction scheme allows to express tensor integrals in terms
of a basis of (4-2$\epsilon$)-dimensional scalar 1-, 2-, 3- and 4-point integrals
 with kinematic coefficients \cite{Brown:1952eu,tHooft:1978xw,Passarino:1978jh}.
While it works well for processes with up to 4 external states, for 5 and more legs the
numerical stability of the reduction coefficients is spoiled by the appearance of inverse Gram determinants.
A number of methods have been proposed to avoid inverse Gram determinants and to improve
the numerical stability~\cite{vanNeerven:1983vr,vanOldenborgh:1989wn,Campbell:1996zw,Binoth:1999sp,Giele:2004iy,Denner:2005nn,%
Binoth:2005ff,Diakonidis:2008ij,Fleischer:2010sq}.

In this contribution we introduce the open-source tensor reduction package PJFry
 which avoids Gram determinant instability
problems by using dimensional recurrence-based algorithms
developed in~\cite{Yundin:phd,AlcarazMaestre:2012vp,Fleischer:2010sq,Denner:2005nn,Binoth:2005ff,Diakonidis:2008ij,Bern:1993kr,Davydychev:1991va,Tarasov:1996br,%
Fleischer:1999hq,Diakonidis:2008dt,Diakonidis:2009fx}.

\section{The reduction algorithm}
We define dimensionally regulated n-point 1-loop tensor integral of rank R as
\begin{align}\label{eq:dimint}
& I_n^{\mu_1\dotsc\mu_R}=(2\pi\mu)^{2\epsilon}\int \frac{d^d k}{i \pi^{d/2}}
  \frac{k^{\mu_1}\dotsm k^{\mu_R}}
{\left((k{-}q_1)^2-m_1^2+i\epsilon\right)\dotsm\left((k{-}q_n)^2-m_n^2+i\epsilon\right)}
\shortintertext{where the chords $q_i$ are defined by the external momenta, see Fig.~\ref{fig:loopplus}:}
&q_1=p_1,\;q_2=p_1+p_2,\;q_3=p_1+p_2+p_3,\;\dotsc,\; q_n=\sum_{i=1}^{n}p_i
\end{align}

Any one loop tensor integral can be rewritten
in terms of scalar tensor form-factors separating the Lorentz structure
 into products of external momenta and metric tensors:
\begin{gather}\label{eq:formfactor}
 I_n^{\mu_1\dotsc\mu_R}=
\sum_{i_1,\dotsc,i_R}^{n} q_{i_1}^{[\mu_1}\dotsm q_{i_R}^{\mu_R]} F^{(n)}_{i_1\dotsc i_R}
+\sum_{i_3,\dotsc,i_R}^{n} {g\mathstrut}^{[\mu_1\mu_2}q_{i_3}^{\mu_3}\dotsm q_{i_R}^{\mu_R]} F^{(n)}_{00i_3\dotsc i_R}
 + \dotsb
\end{gather}
where square brackets denote non-equivalent symmetrization, which gives the set of all
non-equivalent permutations.

Following Ref.~\cite{Davydychev:1991va} tensor form-factors can be mapped to scalar
 integrals in higher dimension and shifted powers of denominators:
\begin{align}\label{eq:davydychev}
 F^{(n)}_{i}    &=-          I_{n,i}^{[2]},
& 
 F^{(n)}_{ij}   &= n_{ij}    I_{n,ij}^{[4]},
&
 F^{(n)}_{00}   &= -\frac12  I_{n}^{[2]},
& 
 F^{(n)}_{ijk}  &=-n_{ijk}   I_{n,ijk}^{[6]},
&
 F^{(n)}_{00k}  &=  \frac12  I_{n,k}^{[4]}.
\end{align}
where $I_{n,i_1i_2\dotsc}^{[2l],s_1s_2\dotsc}$ is a generalized scalar loop integral in shifted dimension:
\begin{gather}\label{eq:hdimint}
 I_{n,i_1i_2\dotsc}^{[2l],s_1s_2\dotsc} =(2\pi\mu)^{2\epsilon} \int \frac{d^{d+2l}k}{i \pi^{d/2+l}} \prod_{r=1}^{n}
\frac{1}{\left((k-q_r)^2-m_r^2+i\epsilon\right)^{1+\delta_{ri_1}+\delta_{ri_2}+\dotsb-\delta_{rs_1}-\delta_{rs_2}-\dotsb}}
\end{gather}
The symbols $n_{i_1i_2\dotsc}$ in \eqref{eq:davydychev} are shorthand notations for combinatorial factors
introduced in \cite{Diakonidis:2008ij}.

Now one can use dimensional recurrence relations derived in Ref.~\cite{Tarasov:1996br,Fleischer:1999hq}
 to express tensor form-factors in terms of (4-2$\epsilon$)-dimensional scalar integrals:
\begin{align}
\label{eq:dimrec}
 ()_n (d+2l-\sum_{i=1}^n\nu_i+1)I_n^{[2(l+1)]}&=
  \mbinom{0}{0}_{\!n} I_n^{[2l]}-\sum_{s=1}^{n}\mbinom{0}{s}_{\!n} \mathbf{s}^- I_n^{[2l]}
\\
\label{eq:combrec}
 ()_n v_j\: \mathbf{j}^+ I_n^{[2(l+1)]}&=
  -\mbinom{j}{0}_{\!n} I_n^{[2l]}+\sum_{s=1}^{n}\mbinom{j}{s}_{\!n} \mathbf{s}^- I_n^{[2l]}
\end{align}
where $\mathbf{s}^-$ is $s^\text{th}$ denominator power lowering operator,
$\mathbf{j}^+$ is $j^\text{th}$ denominator power raising operator and
$\nu_j$ is the power of the $j^\text{th}$ denominator.

However the direct application of eqs.~\eqref{eq:dimrec} and \eqref{eq:combrec} introduces
inverse Gram determinants $()_n$ which will spoil numerical accuracy in large regions of the physical phase
space.

The problem of inverse 5-point Gram determinants has been addressed, among others, in \cite{Fleischer:2010sq}
 where expressions for tensor pentagons have been derived using signed minor algebraic relations.

\begin{figure}[b]
    \centering
    \includegraphics{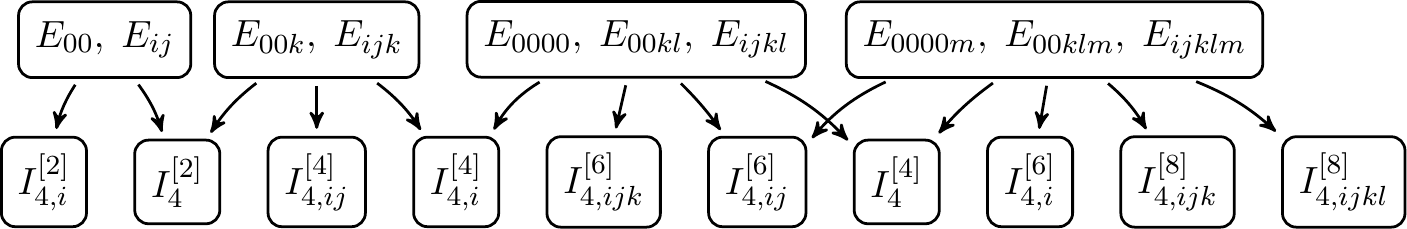}%
    \caption{Basis integrals for the tensor reduction of pentagons}%
    \label{fig:pentbasis}
\end{figure}

The basis shown in Fig.~\ref{fig:pentbasis} can be further reduced using an
 additional recurrence relation:
\begin{align}
\label{eq:indexrec}
\mbinom{0}{0}_{\!n} v_j\: \mathbf{j}^+ I_n^{[2(l+1)]}&=
-\mbinom{j}{0}_{\!n} (d+2l-\sum_{i=1}^n\nu_i+1)I_n^{[2(l+1)]}
+\sum_{s=1}^{n}\mbinom{0s}{0j}_{\!n} \mathbf{s}^- I_n^{[2l]}.
\end{align}
With the help of the above relation one can reduce powers of all denominators to~$1$
without introduction of inverse Gram determinants,
however it does not reduce the dimension.
Thus we cannot recurse all the way down to standard (4-2$\epsilon$)-dimensional scalar integrals
and we have to extend our integral basis:
\begin{align}
\label{eq:highbasis}
   &&I_4^{[8],s},&& I_4^{[6],s},&& I_4^{[4],s},&& I_4^{[2],s},
   && I_3^{[8],st},&& I_3^{[6],st},&& I_3^{[4],st},&& I_3^{[2],st}.
\end{align}

For the numerical evaluation of these additional integrals~\eqref{eq:highbasis}
we employ a series expansion in the small Gram region~\cite{Fleischer:2010sq}.
This method treats all mass-kinematic combinations uniformly with dimensional recurrence, therefore works
equally well for both massive and massless cases.

Our expansion formula is easily derived from \eqref{eq:dimrec}, which can be written in the following form:
\begin{gather}
\label{eq:expand}
{X}(d+2l-n+1) I_n^{[2(l+1)]}= I_n^{[2l]}-Z_n^{[2l]},
\qquad
 {X}=\frac{()_n}{\binom{0}{0}_n},
\qquad
Z_n^{[2l]}=\sum_{s=1}^n\frac{\binom{s}{0}_n}{\binom{0}{0}_n} I_n^{[2l],s}.
\shortintertext{solving the recurrence we get}
{I_n^{[2l]}}=\sum_{m=0}^M a_m^{(l)} {X^m} {Z_n^{[2(l+m)]}}
 + \Bigl[
  a_M^{(l)} X^M I_n^{[2(l+M)}
  \Bigr],
\qquad
a_m^{(l)}=2^m
 \left(\!\frac{d+2l-n+1}{2}\!\right)_{\!m}
\label{eq:expandsolved}
\end{gather}
where $(a)_m=\Gamma(m+a)/\Gamma(a)$ is the Pochhammer symbol.

\section{The PJFry reduction library}
The PJFry package~\cite{Yundin:phd,AlcarazMaestre:2012vp} is an open source library for the numerical evaluation of one loop
tensor integrals. Is is licensed under the GNU Lesser General Public License.
The latest version can be obtained from the program webpage at \href{https://github.com/Vayu/PJFry/}{\texttt{https://github.com/Vayu/PJFry/}}.

The program needs an external library for the evaluation of (4-2$\epsilon$)-dimensional scalar integrals.
Currently QCDLoop \cite{Ellis:2007qk} and OneLOop \cite{vanHameren:2010cp} are supported.
The tensor reduction formulae of the PJFry package rely essentially on Ref. \cite{Fleischer:2010sq}, where
the dimensional recurrence algorithms developed in Refs.~\cite{Tarasov:1996br,Fleischer:1999hq} were explored.
Due to the recursive nature of the numerical algorithms in PJFry, one can greatly benefit from reusing
building blocks throughout the calculation.

The main features of the current public version are:
\begin{itemize}
\item Reduction of up to rank~5 pentagon tensor integrals
\item Any combination of real internal or external masses
\item Leading Gram determinants are avoided by the reduction procedure
\item Subleading small Gram determinants are treated with asymptotic expansions
\item A cache system gives lower point tensor integrals at no extra cost
\item Interfaces for C, FORTRAN, C++, Mathematica and \texttt{GoSam}~\cite{Cullen:2011ac}.
\end{itemize}

\begin{figure}[t]
    \centering
    \includegraphics{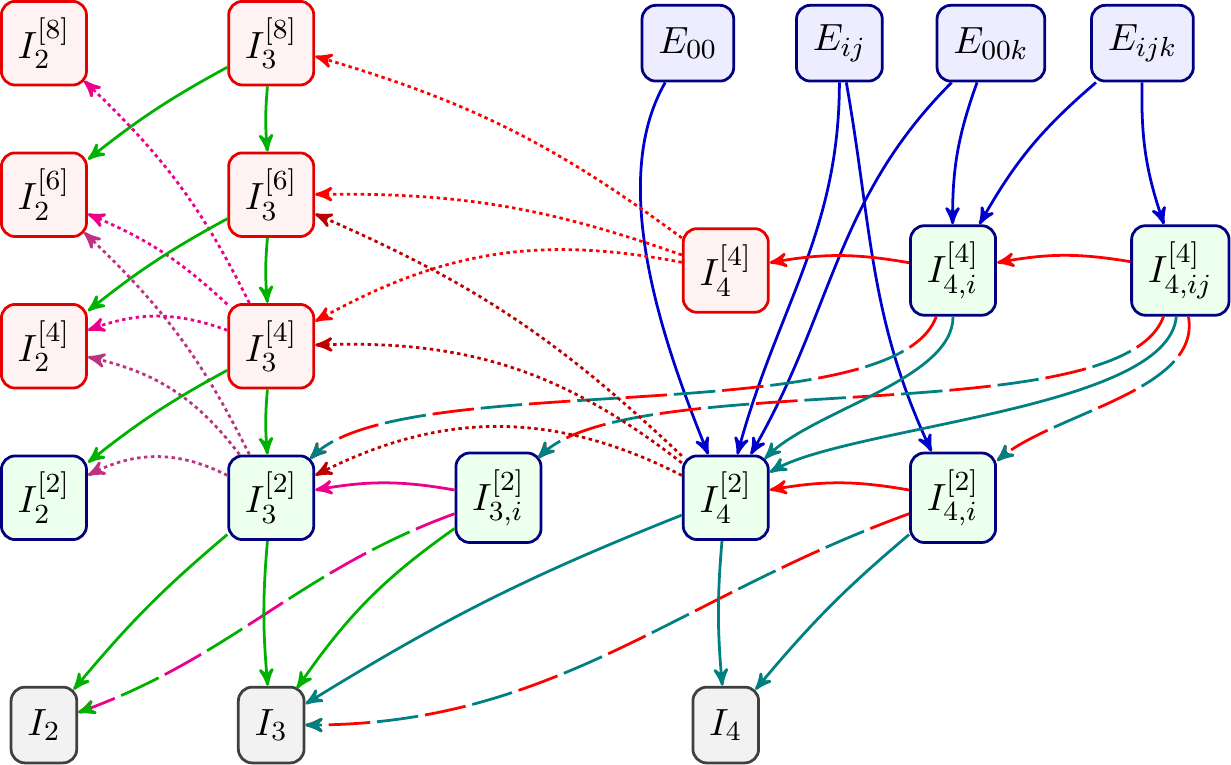}%
    \caption{Rank~2 and~3 pentagon form-factors calculation flowchart}%
    \label{fig:rank3flow}
\end{figure}

The different algorithms used in the program are schematically illustrated for
tensor pentagon form factors of ranks~2 and~3 in Fig.~\ref{fig:rank3flow}.
The blue lines correspond to the reduction formulae of~\cite{Fleischer:2010sq}.
In the reduction of 4-point functions teal lines represent direct downward recursions
for large Gram determinant \eqref{eq:dimrec},
while the solid red paths represent an alternative scheme with Cayley determinants \eqref{eq:indexrec}.
Relations which are shared by both schemes are drawn with two-colored dashed lines.
The small Gram series expansion procedure \eqref{eq:expandsolved} is depicted for boxes by dotted red lines.
Similarly for 3-point functions we use green lines for \eqref{eq:dimrec},
 solid magenta lines for \eqref{eq:indexrec} of $I_{3,i}^{[2]}$ and dotted magenta lines for
the series expansions of scalar $I_3^{[2l]}$ in the small Gram case.


Due to the 4-dimensionality of space-time the problem of inverse Gram determinants does not appear beyond the 
5-point case. Therefore after establishing stable numerical methods for the evaluation of pentagons we can
use them to construct higher point functions without additional effort.

In PJFry we implemented a well known expression~\cite{Denner:2005nn,Binoth:2005ff,Fleischer:1999hq,Diakonidis:2008dt,Diakonidis:2009fx,Diakonidis:2010rs}
 for tensor hexagons up to rank~6 written in terms of pentagons up to rank~5:
\begin{gather}
I_6^{\mu_1\ldots\mu_{R-1}\mu_R} = - \sum^6_{s=1} I_5^{\mu_1\ldots\mu_{R-1},s}
\sum_{i=1}^6 q_i^{\mu_R} \frac{\binom{0s}{0i}_6}{\binom{0}{0}_6}
\end{gather}
Similar representations exist also for 7 and 8-point functions \cite{Fleischer:2011hc,Fleischer:2012hg}.
For higher point functions one might consider using optimized expressions based on contracted tensors
\cite{Fleischer:2011nt}.

In addition to six point functions the upcoming version of PJFry supports complex internal masses
 and extended precision via the qd library~\cite{Bailey:QD}. Both options are available only if using OneLOop~\cite{vanHameren:2010cp}.

The installation is done via a \texttt{configure} script. See the INSTALL file for a detailed description
of available options.

\begin{figure}[t]%
    \centering
    \includegraphics[width=0.8\textwidth,trim=-25pt 0 0 20pt,clip=true]{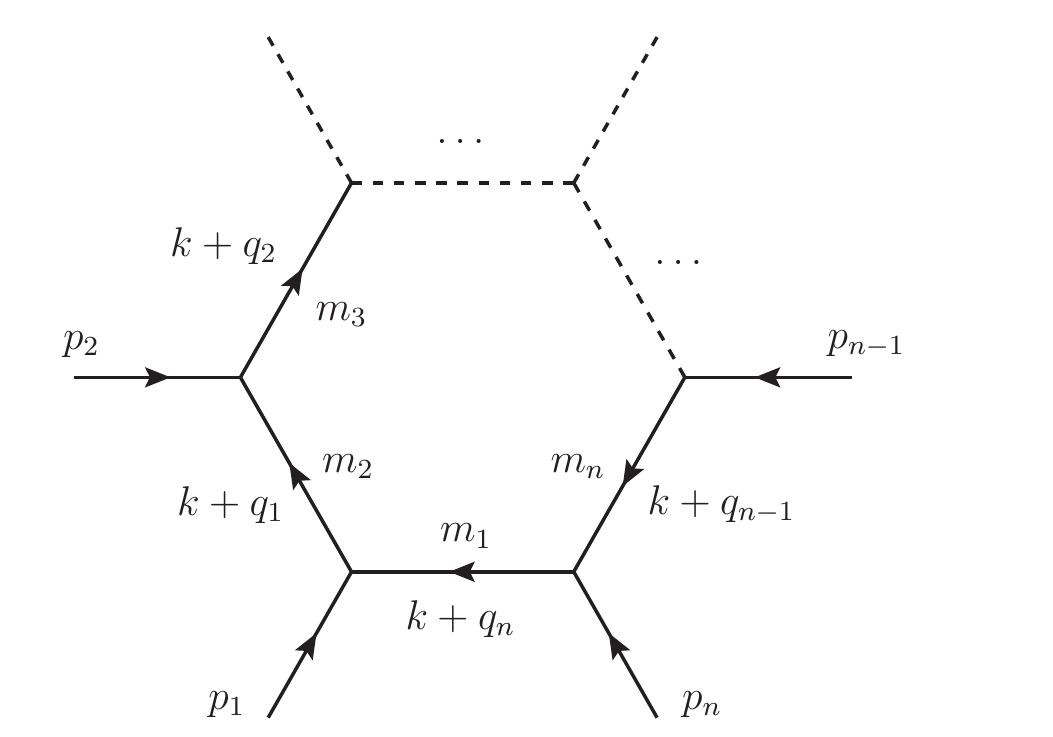}%
    \caption{Momenta labeling}%
    \label{fig:loopplus}
\end{figure}

The Mathematica interface follows the conventions of LoopTools~\cite{Hahn:1998yk} (see
Fig.~\ref{fig:loopplus})
 and can be called from a Mathematica session:
\begin{verbatim}
In:=   Install["PJFry"]
Out:=  PJFry MathLink
Out:=  Type Names["PJFry`*"] to show exported names
\end{verbatim}
The list of exported functions can be obtained by typing
\begin{verbatim}
In:=   Names["PJFry`*"]
Out:=  {"A0v0", "B0v0", "B0v1", "B0v2", "C0v0", "C0v1", "C0v2",
        "C0v3", "ClearCache", "D0v0", "D0v1", "D0v2", "D0v3",
        "D0v4", "E0v0", "E0v1", "E0v2", "E0v3", "E0v4", "E0v5",
        "F0v0", "F0v1", "F0v2", "F0v3", "F0v4", "F0v5", "F0v6",
        "GetMu2", "SetMu2"}
\end{verbatim}
All functions have a short help message which can be accessed by prepending the function
 name in a call with a question mark (e.g. \verb|?F0v1|).

As an example we demonstrate a hexagon tensor form-factor evaluation for one selected phase-space point shown
in Table~\ref{tab1}.

\begin{table}[h]
\begin{center}
\begin{tabular}{|l|llll|}
\hline
$p_1$ & 2.1774554 & 0 & 0 & 2.1774554 \\
$p_2$ & 2.1774554 & 0 & 0 & -2.1774554 \\
$p_3$ & -2.0369414560538   & -0.4757951224 &  0.4212682252 & 0.8409718065   \\
$p_4$ & -2.0907236589150   &  0.5521596147 & -0.466920343  & -0.9001008672  \\
$p_5$ & -0.068463307640486 &  0.0530631952 & 0.02969826743 & -0.03145687079 \\
$p_6$ & -0.15878243799973  & -0.1294276875 & 0.01595385037 & 0.09058593149  \\
\hline
\multicolumn{5}{|c|}{$m_1=0,\quad m_2=0,\quad m_3=0,\quad m_4=3.038049,\quad m_5=0,\quad m_6=0$}\\
\hline
\end{tabular}
\caption{\label{tab1}An example phase-space point.}
\end{center}
\end{table}

\medskip

{\bfseries The scalar hexagon $F_0$:}
{\small
\begin{verbatim}
In:=  k6 = Sequence @@ {ps1, ps2, ps3, ps4, ps5, ps6, s12, s23, s34,
           s45, s56, s16, s234, s345, s456, m1, m2, m3, m4, m5, m6};
In:=  Table[F0v0[k6, ep], {ep, 0, 2}]
Out:= {-30.3579 - 205.8213 I, -48.17500 - 42.76719 I, -10.58591}
\end{verbatim}
}

\medskip

{\bfseries The vector hexagon components $F^\mu$:}
{\small
\begin{verbatim}
In:=  Transpose[Table[
        q1 F0v1[1, k6, ep] + q2 F0v1[2, k6, ep] + q3 F0v1[3, k6, ep]
      + q4 F0v1[4, k6, ep] + q5 F0v1[5, k6, ep], {ep, 0, 2}]]
Out:= {{4.76115 + 38.31618 I, 8.400862 + 7.559091 I, 1.748247},
       {5.93303 + 20.79025 I, 5.326493 + 4.122928 I, 1.080011},
       {0.162103 - 6.809284 I, -1.119560 - 1.201034 I, -0.2433933},
       {-4.02995 - 15.03100 I, -3.642480 - 3.047085 I, -0.7563652}}
\end{verbatim}
}

\section{Acknowledgments}
V.\,Y. would like to thank the organizers of Loops and Legs 2012
 for their warm hospitality and a perfectly organized conference.

\providecommand{\href}[2]{#2}
\addcontentsline{toc}{section}{References}
\bibliographystyle{JHEP}
\bibliography{bibliography}

\end{document}